\documentclass[conference]{IEEEtran}
\IEEEoverridecommandlockouts

\usepackage{cite}
\usepackage{amsmath,amssymb,amsfonts}
\usepackage{graphicx}
\usepackage[caption=false,font=footnotesize]{subfig} 
\usepackage{textcomp}
\usepackage{xcolor}
\usepackage{url}
\usepackage{algorithm}
\usepackage{algpseudocode} 

\def\BibTeX{{\rm B\kern-.05em{\sc i\kern-.025em b}\kern-.08em
    T\kern-.1667em\lower.7ex\hbox{E}\kern-.125emX}}

\begin{document}

\title{Data Scarcity in Gas Load Profiling: Generalized Proxy-Guided Load and Temporal Disaggregation}

\author{
  \IEEEauthorblockN{1\textsuperscript{st} Lucas Krome}
  \IEEEauthorblockA{\textit{Dept. of Electrical, Computer,} \\
    \textit{and Biomedical Engineering} \\
    \textit{Toronto Metropolitan University}\\
    Toronto, ON, Canada \\
    lucas.krome@torontomu.ca}
  \and
  \IEEEauthorblockN{2\textsuperscript{nd} Mahtab Aboufazeli}
  \IEEEauthorblockA{\textit{Dept. of Electrical, Computer,} \\
    \textit{and Biomedical Engineering} \\
    \textit{Toronto Metropolitan University}\\
    Toronto, ON, Canada \\
    mahtab.aboufazeli@torontomu.ca}
  \and
  \IEEEauthorblockN{3\textsuperscript{rd} Soosan Beheshti}
  \IEEEauthorblockA{\textit{Dept. of Electrical, Computer,} \\
    \textit{and Biomedical Engineering} \\
    \textit{Toronto Metropolitan University}\\
    Toronto, ON, Canada \\
    soosan@torontomu.ca}
  \and
  \IEEEauthorblockN{4\textsuperscript{th} Bala Venkatesh}
  \IEEEauthorblockA{\textit{Dept. of Electrical, Computer,} \\
    \textit{and Biomedical Engineering} \\
    \textit{Toronto Metropolitan University}\\
    Toronto, ON, Canada\\
    bala@torontomu.ca}
}

\author{
\IEEEauthorblockN{Lucas Krome}
\IEEEauthorblockA{
\textit{Dept. of Electrical, Computer, and Biomedical Engineering}\\
Toronto Metropolitan University\\
Toronto, ON, Canada\\
lucas.krome@torontomu.ca}
\and

\IEEEauthorblockN{Mahtab Aboufazeli}
\IEEEauthorblockA{
\textit{Dept. of Electrical, Computer, and Biomedical Engineering}\\
Toronto Metropolitan University\\
Toronto, ON, Canada\\
mahtab.aboufazeli@torontomu.ca}
\and

\IEEEauthorblockN{Soosan Beheshti}
\IEEEauthorblockA{
\textit{Dept. of Electrical, Computer, and Biomedical Engineering}\\
Toronto Metropolitan University\\
Toronto, ON, Canada\\
soosan@torontomu.ca}
\and

\IEEEauthorblockN{Bala Venkatesh}
\IEEEauthorblockA{
\textit{Dept. of Electrical, Computer, and Biomedical Engineering}\\
Toronto Metropolitan University\\
Toronto, ON, Canada\\
bala@torontomu.ca}
}

\maketitle

\begin{abstract}
The electrification of heating systems is a critical pathway for decarbonizing the building sector; however, the development of sustainable strategies is often hindered by the lack of granular thermal load profiles. The nature of this problem is such that available data are extremely scarce, irregular, and low-frequency, rendering conventional data-hungry machine learning approaches impractical or unreliable. High-resolution gas metering is rarely available, as it falls outside standard utility business requirements. To bridge this gap and enable data-driven AI analytics under severe data limitations, this paper presents the Generalized Proxy-Guided Load and Temporal Disaggregation framework. Validated on a dataset of 11 multi-unit residential buildings over 18 months, the framework achieves a mean squared percentage error (MSPE) of 6.37\% for reconstructed total gas consumption. The methodology operates through a four-stage process: (i) Generalized Occupancy Proxy Extraction via weather normalization to isolate behavioral signals from hourly electricity data; (ii) Unified Segmentation and Normalized Pooling to mitigate the statistical limitations of sparse billing data; (iii) Unified Baseload Parameter Estimation enhanced by local calibration to ensure building-specific accuracy; and (iv) Component-Wise Temporal Disaggregation to reconstruct distinct baseload and heating profiles. Overall, the proposed framework effectively bridges the resolution gap, transforming low-frequency utility data into high-fidelity training sets suitable for scalable, data-driven decarbonization modeling.
\end{abstract}


\section{Introduction}
Decarbonizing Canada's building sector which is responsible for 18\% of national emissions is critical for 2050 net-zero goals. While heat pumps are essential, they require precise thermal load profiles often unavailable in older stocks. Advanced forecasting models, from "black-box" DNNs \cite{wang2020building, geysen2018operational, leiprecht2021comprehensive} to "white-box" tools like EnergyPlus \cite{crawley2001energyplus} and hybrid approaches \cite{zhao2023artificial}, demand high-resolution data. However, the lack of smart metering renders these inapplicable for retrofits relying on monthly billing \cite{wang2017review}.

To address scarcity, researchers traditionally used inverse modeling like variable-based degree-day methods \cite{fels1986prism} or change-point regression \cite{kissock1998ambient}. Despite advancements in uncertainty quantification \cite{chong2018guidelines} and model selection \cite{granderson2015automated}, these macroscopic tools lack the intra-day resolution required for equipment sizing.

Consequently, research has shifted toward temporal downscaling. Early methods used rectangular shaping \cite{lamagna2020hourly} or simulation benchmarks \cite{fumo2010methodology, smith2011robustness}, though \cite{stegner2019comparing} showed these often fail to capture individual variability. Newer techniques map monthly bills to hourly profiles using k-NN \cite{lazzeroni2023data} or Random Forests \cite{giannuzzo2024reconstructing}, but typically rely on static schedules that miss stochastic occupant behavior \cite{de2016modelling}.

This paper bridges the resolution gap by introducing a framework that reconstructs hourly gas profiles using monthly billing and hourly electricity measurements. Utilizing residual electricity as a temporal occupancy proxy, the algorithm modulates gas baseloads to ensure physical consistency. The paper is organized as follows: Section \ref{sec:data} details the data and motivation; Section \ref{sec:method} presents the methodology; Section \ref{sec:results} validates the framework against residential portfolios and the AMPds dataset; and Section \ref{sec:conclusions} concludes.

\begin{figure*}[t]
    \centering
    \includegraphics[ width=0.8\textwidth]{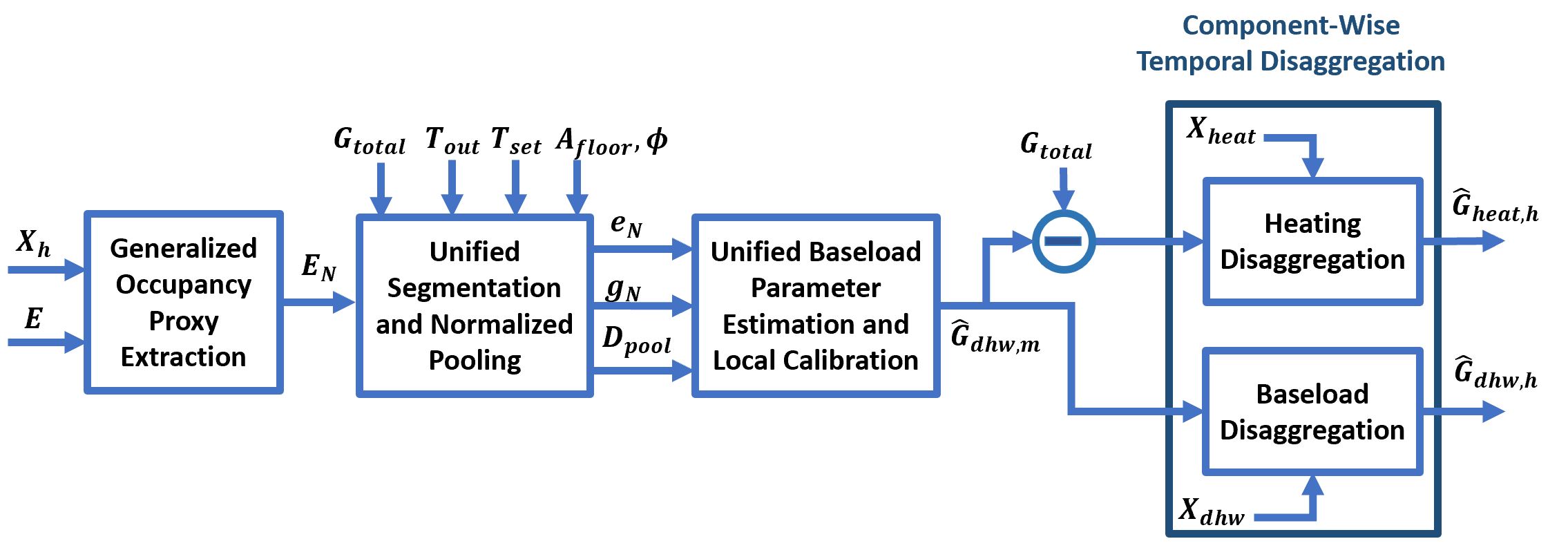}
    \caption{Block diagram of the Generalized Proxy-Guided Load and Temporal Disaggregation algorithm.}
    \label{fig:flowchart}
\end{figure*}

\section{Data Description and Motivation}\label{sec:data}
The framework is driven by three core objectives: (1) Addressing Data Scarcity: Overcoming the lack of sub-metered gas data; (2) Generalizable Proxy Integration: Capturing stochastic behavior using hourly electricity as a dynamic, adaptable proxy; and (3) Data Synthesis: Generating high-fidelity training datasets from low-frequency billing records for AI analytics.

The study utilizes operational data from 11 multi-unit residential buildings over 18 months (Jan 2022–June 2023). The dataset integrates: (1) Utility Data (monthly gas $m^3$, hourly electricity kWh); (2) Meteorological Data (temperature, wind speed, humidity); and (3) Building Characteristics (floor area, volume, orientation). Manual inspection confirmed high data integrity, requiring negligible imputation.

\section{Methodology: Generalized Proxy-Guided Load and Temporal Disaggregation}\label{sec:method}
The presented framework, termed \textit{Generalized Proxy-Guided Load and Temporal Disaggregation}, addresses the challenge of data sparsity in residential gas profiling. It is premised on the physical decomposition of total natural gas consumption ($G_{total}$) into two distinct end-uses: space heating ($G_{heat}$) and domestic hot water ($G_{dhw}$).
\begin{equation}
    G_{total} = G_{heat} + G_{dhw}
\end{equation}

Meteorological conditions govern space heating, whereas domestic hot water is stochastic and driven by occupant behavior \cite{spaceHeating2018}, and constitutes the system's baseload consumption.
The framework proceeds in four sequential steps to decouple these loads and reconstruct high-resolution profiles and the block diagram is displayed in Fig. \ref{fig:flowchart}.

\subsection{Step 1: Generalized Occupancy Proxy Extraction via Weather Normalization}
Accurate baseload modeling requires a dynamic variable to represent the stochastic nature of occupant behavior. We identify hourly electricity consumption as the most accessible high-resolution tracer for this activity. However, raw electricity data ($E$) is often confounded by seasonal cooling loads.
To isolate the latent behavioral signal, we apply Multivariate Weather Normalization using a Multiple Linear Regression (MLR) framework:
\begin{equation}
    E = \underline{\gamma}^\top \mathbf{X}_h + \epsilon
\end{equation}
Here, $\underline{\gamma}$ is the coefficient vector and $\mathbf{X}_h$ is the feature vector comprising meteorological and temporal priors:
\begin{equation}
    \mathbf{X}_h = [1, CDH, W_{s}, H, I_{wknd}, \bar{E}_{hr,loc}, \bar{E}_{mo,glob}]^\top
    \label{eqn:elec_Xh}
\end{equation}
where $W_{s}$ is wind speed (m/s), $H$ is relative humidity (\%), and $I_{wknd}$ is a binary weekend indicator. Cooling Degree Hours ($CDH$) are defined as $CDH = \max(T_{out} - T_{cool}, 0)$, capturing thermal AC demand. 

In (\ref{eqn:elec_Xh}), the terms $\bar{E}_{hr,loc}$ and $\bar{E}_{mo,glob}$ denote temporal priors designed to isolate recurring behavioral routines: (1) Local Hourly Profile ($\bar{E}_{hr,loc}$): The building-specific mean consumption for each hour ($0\text{--}23$ h), capturing unique diurnal peak timings; (2) Global Monthly Profile ($\bar{E}_{mo,glob}$): The ensemble average across all buildings, capturing the shared seasonal intensity baseline of the portfolio.

The final Occupancy Proxy ($E_N$) is extracted by setting the thermal components ($CDH, W_s, H$) to zero in the  model:
\begin{equation}
    E_N = \max(\mathbf{\gamma}^\top \mathbf{X}'_h, 0)
\end{equation}
where $\mathbf{X}'_h = [1, 0, 0, 0, I_{wknd}, \bar{E}_{hr,loc}, \bar{E}_{mo,glob}]^\top$. This formulation yields a clean behavioral signal, preserving usage patterns driven by lighting and appliances while removing cooling spikes.

\subsection{Step 2: Unified Segmentation and Normalized Pooling}
To isolate the baseload consumption ($G_{dhw}$), we first perform a thermodynamic segmentation to identify the "non-heating" period where $G_{heat} \approx 0$. This study defines the non-heating season as any month where the daily average Heating Degree Hours (HDH) is less than 5. HDH is calculated by
\begin{equation}
    HDH = \max(T_{set}-T_{out},0)
\end{equation}
where $T_{out}$ and $T_{set}$ represent the outdoor temperature and operational setpoint temperature, respectively.
To mitigate the statistical insignificance of sparse individual records, we employ a Unified Pooling Strategy. Observations from the identified non-heating months across all 11 buildings are structurally normalized to create comparable intensity metrics. Electricity is normalized by floor area ($e_N = E_N / A_{floor}$) to reflect plug-load density, while gas consumption is normalized by the specific volume-to-area ratio of the building ($g_{N} = G_{total} / \phi$) where $\phi = V / A_{floor}$ and $V$ represents the volume of the building. Similarly $g_{dhw} = G_{dhw}/\phi$. These normalized vectors are then aggregated into a unified dataset $\mathcal{D}_{pool}$, allowing the extraction of robust trends that are invisible at the individual building level.

\subsection{Step 3: Unified Baseload Parameter Estimation and Local Calibration}
With the normalized proxy ($e_N$) validated, we formulate the baseload model using a Unified Baseload Regression.
A generalized Ordinary Least Squares (OLS) model is trained on the pooled dataset $\mathcal{D}_{pool}$ to learn the global sensitivity of base usage to occupant behavior. The standardized model is given by:
\begin{equation}
    g_{dhw} = \beta_0 + \beta_1 e_N + \beta_2 \bar{e}_{hr} + \beta_3 \bar{e}_{mo} + \epsilon
\end{equation}
where $\beta_{0..3}$ are the unified coefficients representing the baseline intensity and the marginal impact of occupant activity common across the stock.

The complete reconstruction process is detailed in Algorithm \ref{alg:baseload_est}. Following the extraction of unified parameters, the model applies a Local Calibration step. To account for building-specific variations in appliance efficiency, the unified prediction is adjusted via a scalar calibration factor $\kappa_i$. This factor is derived by minimizing the residual between the universal prediction and the observed gas consumption during the building's most representative non-heating month.

\begin{algorithm}[htbp]
\caption{Unified Baseload Disaggregation Framework}
\label{alg:baseload_est}
\begin{algorithmic}
\renewcommand{\alglinenumber}[1]{} 
\Require $E, G_{total}, T_{out}, W_s, H, A_{floor}, V, T_{set}, T_{cool}$
\Ensure $G_{dhw}$

\Statex \textbf{1. Proxy Extraction:} $\forall i$:
\State \quad $\underline{\gamma}_i \gets \arg\min_{\gamma} \| E_i - \mathbf{X}_{h,i} \gamma \|^2$
\State \quad $E_{N,i} \gets \max(\underline{\gamma}_i^\top \mathbf{X}'_{h,i}, 0)$ 

\Statex \textbf{2. Pooling:} Initialize $\mathcal{D}_{pool} \gets \emptyset$. For all $i, t$ where $\text{mean}(HDH) < 5$:
\State \quad $e_{N} \gets E_{N,i,t} / A_{floor,i}; \quad g_{N} \gets G_{total,i,t} / \phi_i$
\State \quad $\mathcal{D}_{pool} \gets \mathcal{D}_{pool} \cup \{ (e_N, g_N) \}$

\Statex \textbf{3. Estimation \& Calibration:}
\For{each building $i$}
    \State $\hat{G}_{pre,i} \gets (\boldsymbol{\beta} \cdot [1, e_{N,i}, \bar{e}_{hr}, \bar{e}_{mo}]) \times \phi_i$
    \State $\kappa_i \gets G_{total,i,t^*} / \hat{G}_{pre,i,t^*}$ \\ 
           \quad \quad \textbf{where} $t^* \gets \arg\min |G_{total} - \hat{G}_{pre}|$
    \State $G_{dhw,i} \gets \kappa_i \hat{G}_{pre,i}$
\EndFor
\end{algorithmic}
\end{algorithm}
\subsection{Step 4: Component-Wise Temporal Disaggregation}
To generate high-resolution building performance data, we employ a dual-stream disaggregation framework. This process separates the total monthly gas consumption into two distinct physical regimes \textit{i.e.}, the domestic hot water or the baseload ($G_{dhw}$) and the Heating ($G_{heat}$) and reconstructs their hourly profiles using regime-specific drivers.

First, the monthly heating load is isolated as the residual of the baseload model: $G_{heat,m} = G_{total,m} - G_{dhw,m}$. We then apply an Optimization-Based Temporal Downscaling algorithm separately to each component:

\begin{enumerate}
    \item Baseload Reconstruction: Driven by occupant behavior, modeled using the electrical proxy features ($\mathbf{X}_{dhw} = [1, e_N, \bar{e}_{hr}, \dots]$).
    \item Heating Reconstruction: Driven by thermodynamics, modeled using meteorological features ($\mathbf{X}_{heat} = [HDH, W_{s}, H]$).
\end{enumerate}

For each component $k \in \{dhw, heat\}$, we solve for a coefficient vector $\boldsymbol{\beta}_k$ that minimizes the error between the aggregated hourly predictions and the known monthly targets. The objective function is:
\begin{equation}
    \min_{\boldsymbol{\beta}_k} \sum_{m} \left( G_{k,m} - \sum_{t \in m} \max(\mathbf{X}_{k,t} \cdot \boldsymbol{\beta}_k, 0) \right)^2
\end{equation}
This formulation ensures that the synthesized hourly profiles strictly conserve the energy mass of the monthly bills while adhering to the specific temporal dynamics of their respective physical drivers (e.g., occupancy schedules for hot water vs. cold fronts for heating).
\begin{algorithm}[htbp]
\caption{Dual-Stream Hourly Reconstruction}
\label{alg:dual_disagg}
\begin{algorithmic}
\Require Monthly Targets $G_{dhw}, G_{total}$, Features $\mathbf{X}_{dhw}, \mathbf{X}_{heat}$
\Ensure Hourly Profiles $\mathbf{g}_{dhw}, \mathbf{g}_{heat}$

\item[] \textbf{1. Separation:} Calculate Monthly Heating Residuals
\item[] \quad $G_{heat,m} \leftarrow G_{total,m} - G_{dhw,m}$

\item[] \textbf{2. Optimization Routine:}
\item[] \textbf{function} \textsc{Reconstruct}($G_{target}, \mathbf{X}$)
\item[] \quad Minimize $L(\boldsymbol{\beta}) = \sum_{m} ( G_{target,m} - \sum_{t \in m} \max(\mathbf{X}_{t} \boldsymbol{\beta}, 0) )^2$
\item[] \quad $\boldsymbol{\beta}^* \leftarrow \arg\min_{\beta} L(\beta)$ \hfill \scalebox{0.8}{$\triangleright$ Coefficient estimation}
\item[] \quad \textbf{return} $\max(\mathbf{X} \boldsymbol{\beta}^*, 0)$ \hfill \scalebox{0.8}{$\triangleright$ Reconstructed profile}
\item[] \textbf{end function}

\item[] \textbf{3. Execute Dual Streams:}
\item[] \quad $\mathbf{g}_{dhw} \leftarrow \textsc{Reconstruct}(G_{dhw}, \mathbf{X}_{dhw})$ \hfill \scalebox{0.8}{$\triangleright$ Baseload stream}
\item[] \quad $\mathbf{g}_{heat} \leftarrow \textsc{Reconstruct}(G_{heat}, \mathbf{X}_{heat})$ \hfill \scalebox{0.8}{$\triangleright$ Heating stream}

\item[] \textbf{return} $\mathbf{g}_{dhw}, \mathbf{g}_{heat}$
\end{algorithmic}
\end{algorithm}

\section{Results and Discussion}\label{sec:results}
The framework was evaluated on the 18-month dataset across four stages: characterizing occupancy patterns, validating proxy extraction, estimating unified baseload parameters, and verifying component-wise disaggregation.

\subsection{Characterization of Temporal Occupancy Patterns}\label{sec:results_corr}
\begin{figure}[htbp]
    \centering
    \subfloat[Global Monthly Profile\label{fig:monthly}]{
        \includegraphics[width=0.9\linewidth]{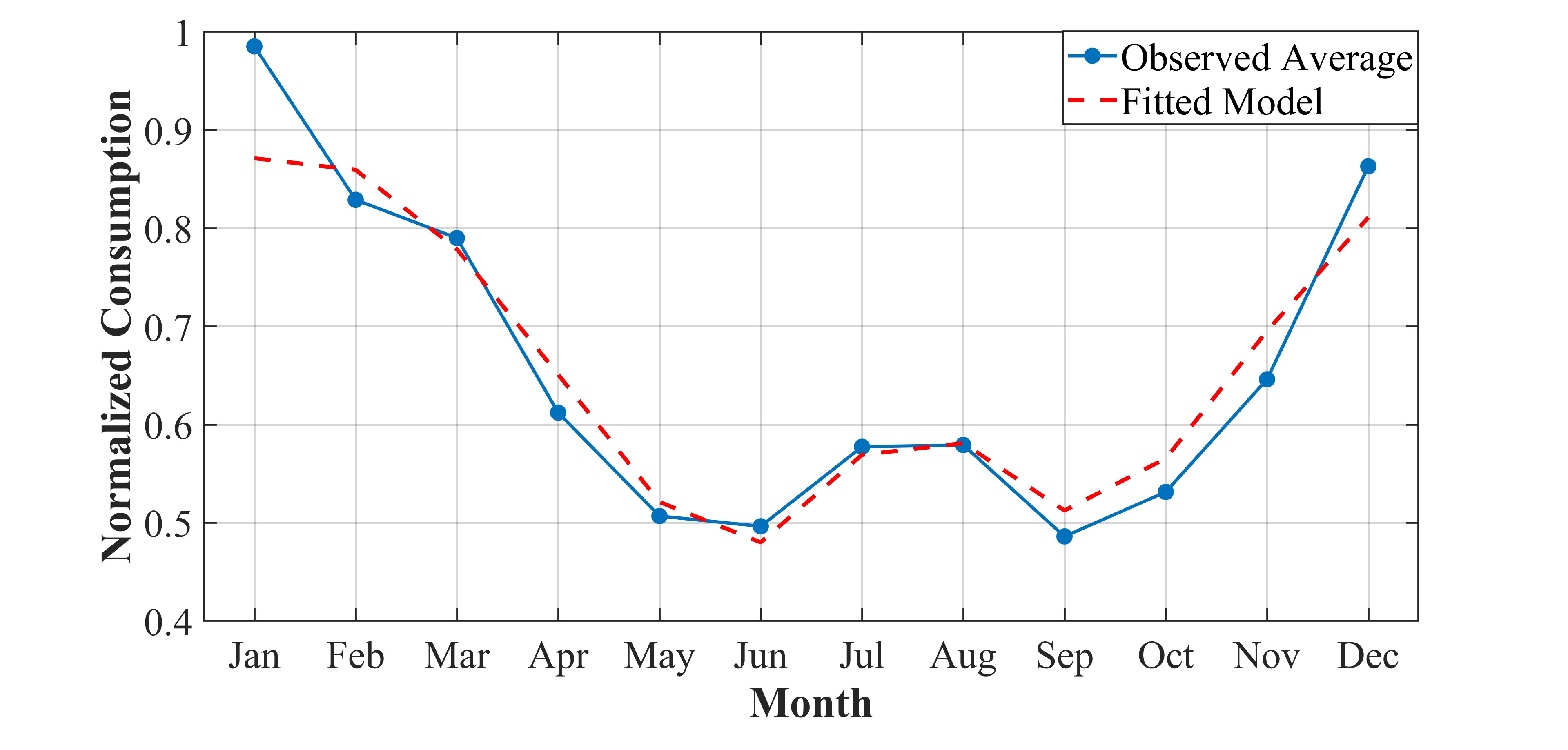}
    }
    \\ 
    \vspace{0.5em}
    \subfloat[Local Hourly Profiles\label{fig:hourly}]{
        \includegraphics[width=0.9\linewidth]{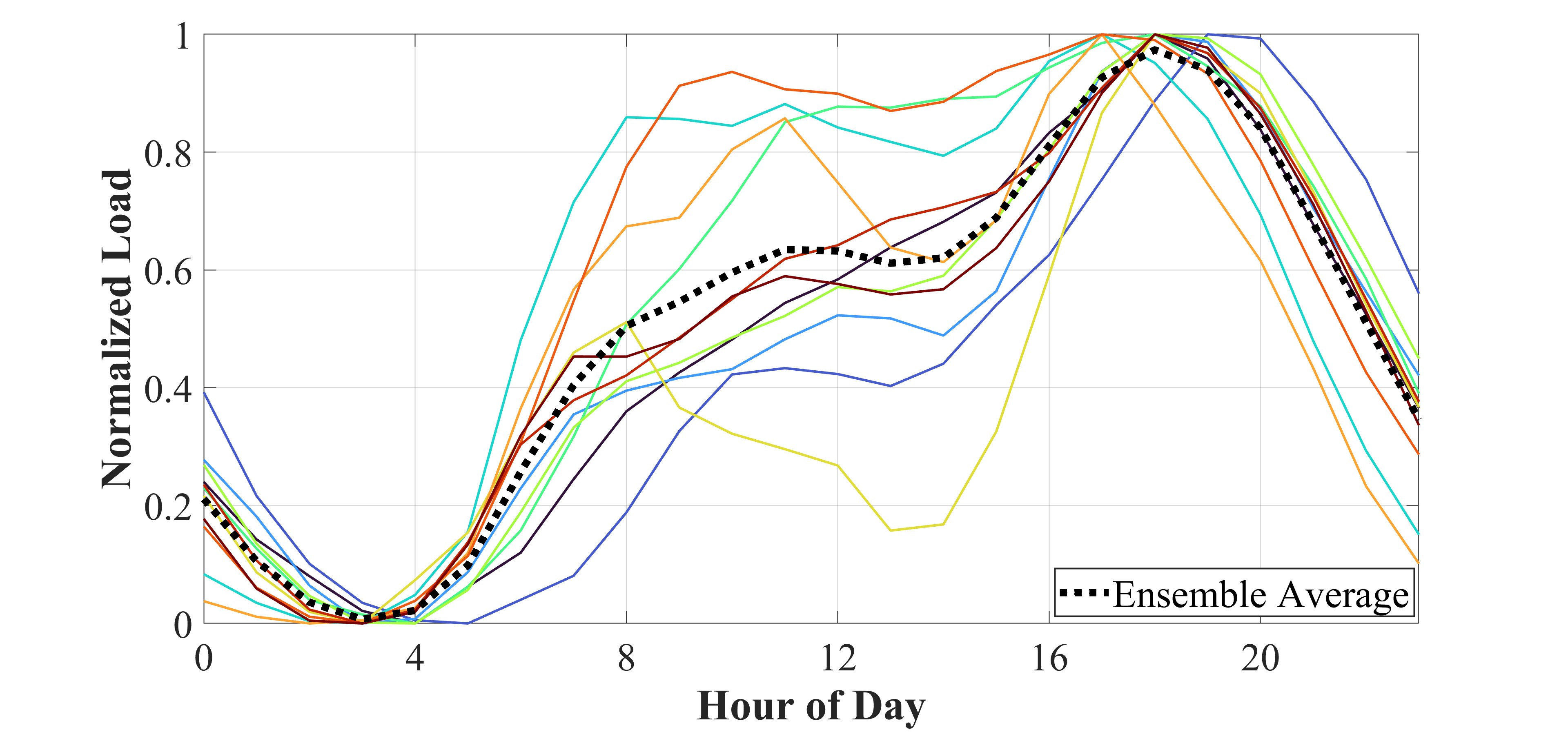}
    }
    \caption{Learned Temporal Characteristics. (a) Global Monthly Profile ($\bar{E}_{mo,glob}$) captures shared baseline seasonality. (b) Local Hourly Profiles ($\bar{E}_{hr,loc}$) reveal high daily variability (solid) compared to the ensemble mean (dotted).}
    \label{fig:temporal_profiles}
\end{figure}

The weather normalization process successfully extracted distinct temporal signatures that characterize the behavioral baseline of the building stock. Figure \ref{fig:temporal_profiles} visualizes the two learned profile components.

Fig. \ref{fig:monthly} displays a Global Monthly Profile that represents a generalized seasonal curve derived by averaging the normalized consumption of all buildings, fitted to a parametric model (sinusoidal + Gaussian). 
The Local Hourly Profiles (Fig. \ref{fig:hourly}) demonstrate the diversity of daily schedules across the 11 buildings. 
While the ensemble average (dotted line) follows a standard dual-peak residential curve, individual buildings exhibit distinct peak timings and base intensities. This confirms the necessity of including building-specific hourly priors ($\bar{E}_{hr,loc}$) rather than relying on a generic standard profile.

\subsection{Validation of the Occupancy Proxy ($E_N$)}\label{sec:results_corr}
To verify that these extracted profiles accurately reflect hot water usage, we performed a comparative correlation analysis against gas consumption during the non-heating season.

\begin{figure}[htbp]
    \centering
    \includegraphics[width=\linewidth]{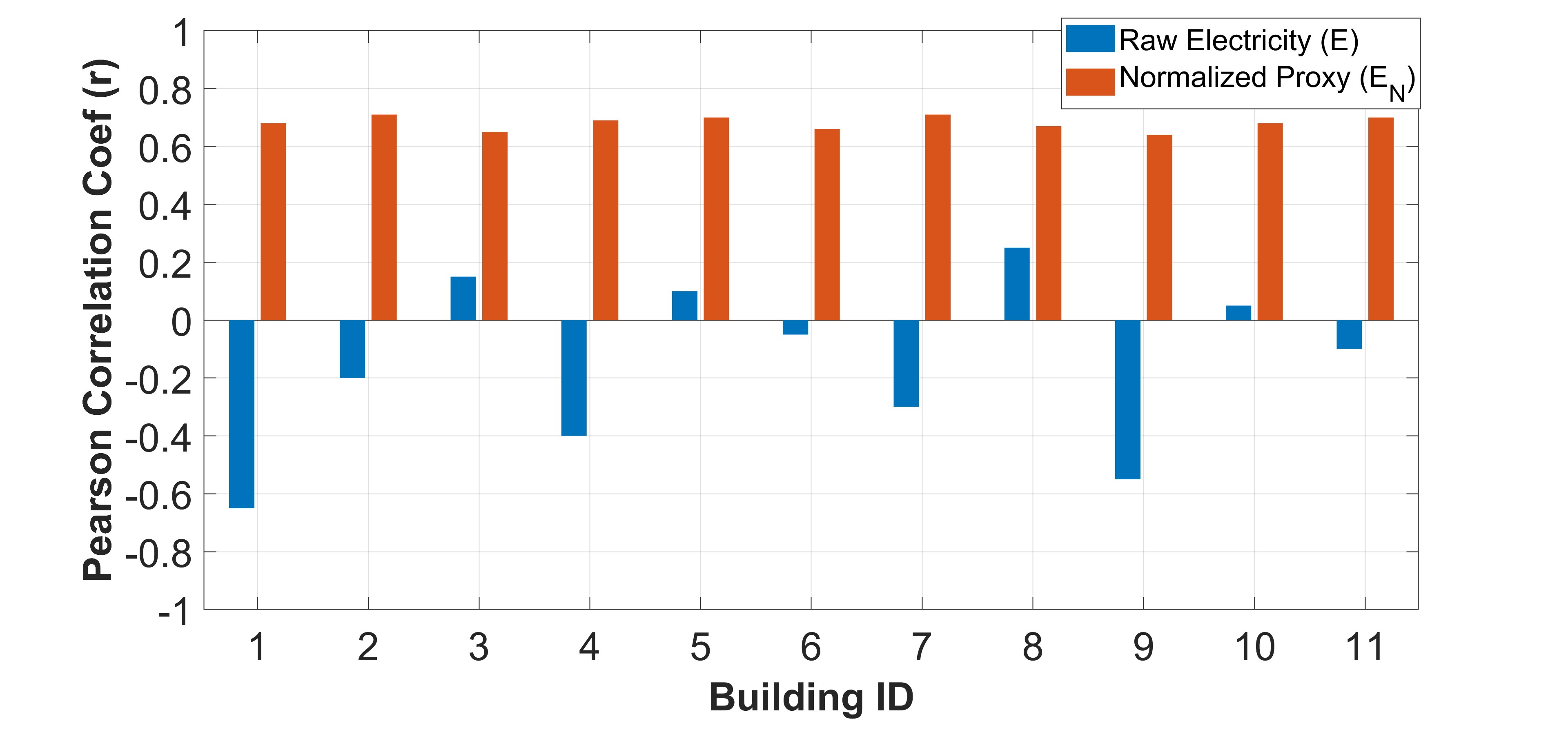}
    \caption{Validation of Occupancy Proxy. Pearson correlation ($r$) of $G_{total}$ with $E$ (blue) vs. $E_N$ (red) across 11 buildings. Red bars demonstrate consistently strong correlation, overcoming the cooling interference evident in the blue bars.}
    \label{fig:correlation_barplot}
\end{figure}

Figure \ref{fig:correlation_barplot} contrasts the performance of Raw Electricity ($E$) against the Normalized Proxy ($E_N$). Raw electricity exhibits weak correlations (mean $\{r\}=0.28$) due to cooling contamination. In contrast, the normalized proxy $E_N$ yields consistently strong positive correlations (mean $\{r\}=0.74$). This systematic improvement confirms that the derived temporal profiles successfully isolate the occupant behavior required for robust baseload modeling.

\subsection{Performance of Unified Baseload Parameter Estimation and Local Calibration}
Next, we evaluate the \textit{Unified Baseload Parameter Estimation and Local Calibration}. This component uses the pooled dataset ($\mathcal{D}_{pool}$) to establish a unified relationship linking temporal and occupancy proxies to the domestic hot water consumption ($G_{dhw}$).

\begin{figure}[htbp]
    \centering
    \includegraphics[width = 0.75\linewidth]{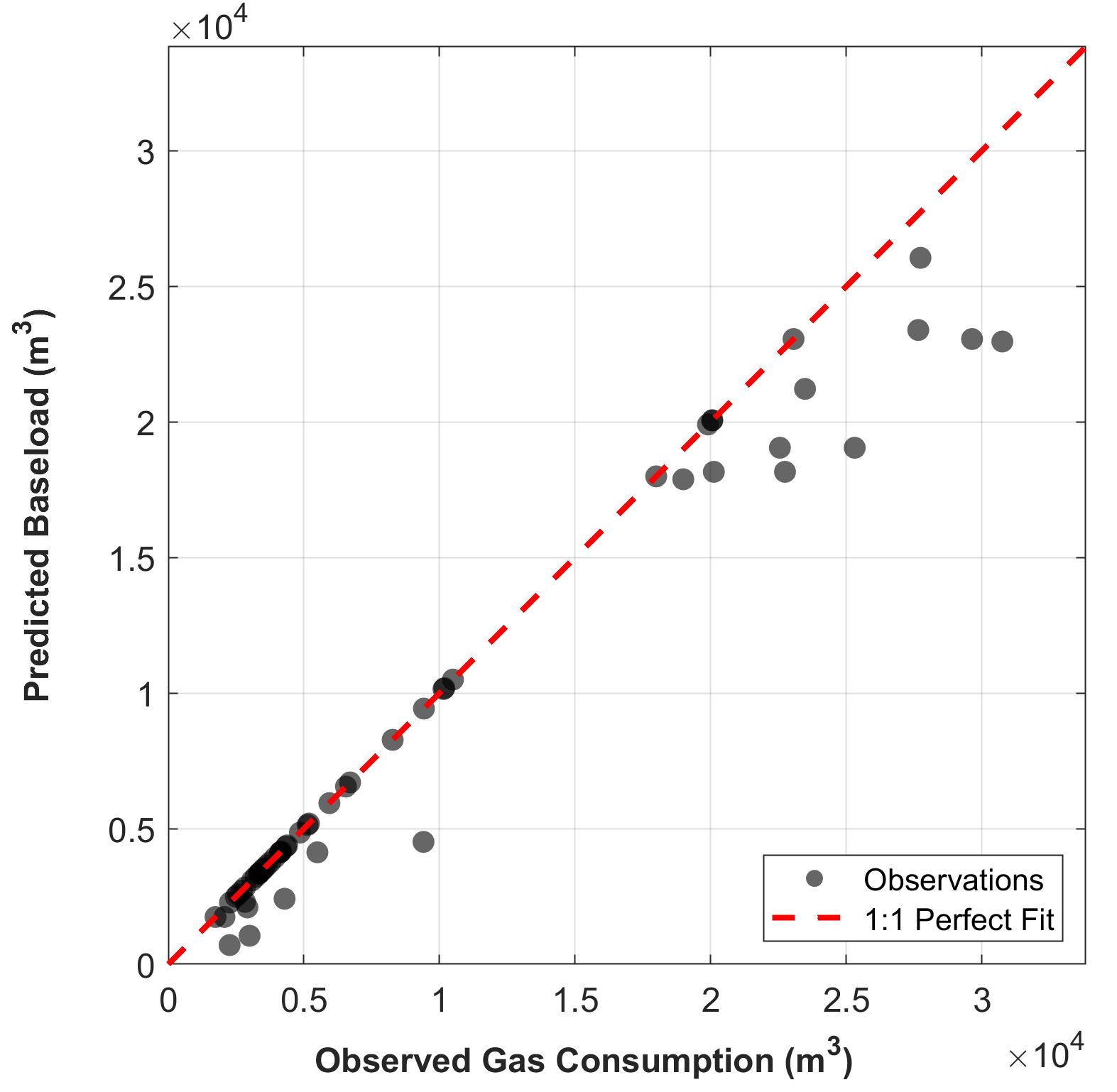}
    \caption{Predicted vs. Observed Gas (non-heating). Tight clustering around the 1:1 line confirms baseload accuracy.}
    \label{fig:baseload_parity}
\end{figure}

The pooled regression captured global occupancy trends, while local calibration ($\kappa_i$) adjusted for building-specific efficiencies. Figure \ref{fig:baseload_parity} confirms the model's accuracy during non-heating months, where tight clustering around the unity line indicates negligible systematic bias

Quantitatively, the model achieved a Mean Squared Percentage Error (MSPE) of 8.58\% and a normalized Root Mean Square Error (CV-RMSE) of 21.65\% across the portfolio for the baseload period. These metrics confirm that the \textit{Unified Pooling} strategy successfully overcomes individual data sparsity, generating a reliable baseload even for buildings with limited historical records.

\subsection{Total Load Reconstruction Validation}

Since sub-metered component data is unavailable for the target buildings, the primary validation for this dataset involves comparing the aggregated model output ($\hat{G}_{total} = \hat{G}_{dhw} + \hat{G}_{heat}$) against the ground-truth total gas consumption ($G_{total}$) recorded by the smart meters. This step ensures that the disaggregation process satisfies the physical energy balance and correctly captures the temporal dynamics of the building.

\begin{figure}[t!]
    \centering
    \includegraphics[width=\linewidth]{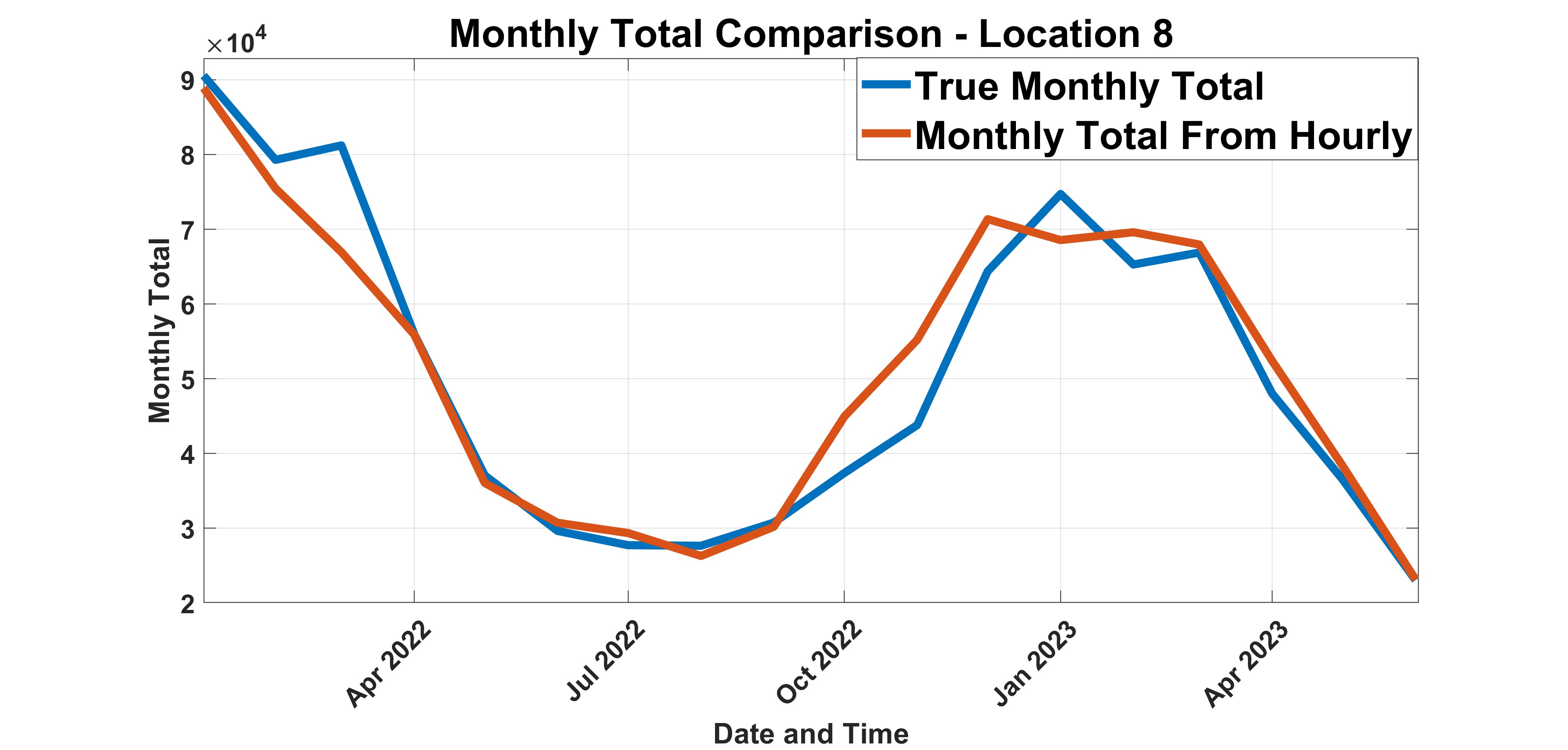}
    \caption{Total Load Reconstruction. Modelled $\hat{G}_{total}$ (red) closely tracks actual $G_{total}$ (blue) for building 8, capturing both high-frequency baseload variability and heating peaks.}
    
    \label{fig:total_load_comparison}
\end{figure}

Figure~\ref{fig:total_load_comparison} presents the time-series comparison for a representative building. The modeled total profile demonstrates strong agreement with the actual consumption, successfully tracking both the high-frequency baseload variations and the magnitude of heating peaks without significant amplitude distortion or temporal lag.

To quantify performance across the entire portfolio, the reconstruction accuracy for all 11 buildings is summarized in Table~\ref{tab:accuracy_metrics}. The model achieves a high degree of consistency, with an average Normalized Root Mean Square Error (NRMSE) of 0.20 and a Mean Absolute Percentage Error (MAPE) of 6.37\%. These results confirm that the dual-stream framework generates a robust total load profile that strictly conserves the billing energy while reflecting the operational reality of the users.

\begin{table}[t!]
    \centering
    \caption{Validation Results. Summary of error statistics for total gas consumption ($G_{total}$) for all 11 buildings.}
    \label{tab:accuracy_metrics}
    \begin{tabular}{|c|c|c|c|}
        \hline
        \textbf{Building ID} & \textbf{NRMSE (-)} & \textbf{RMSE (kWh)} & \textbf{MSPE (\%)} \\
        \hline
         01 & 0.18 & 1109.92 & 2.61 \\
        \hline
         02 & 0.22 & 1681.87 & 6.07 \\
        \hline
         03 & 0.35 & 1866.75 & 37.71 \\
        \hline
         04 & 0.26 & 1445.00 & 4.62 \\
        \hline
         05 & 0.11 & 1319.46 & 1.50 \\
        \hline
         06 & 0.14 & 2103.62 & 0.99 \\
        \hline
         07 & 0.25 & 1205.78 & 5.89 \\
        \hline
         08 & 0.11 & 5833.67 & 1.65 \\
        \hline
         09 & 0.33 & 6795.71 & 5.11 \\
        \hline
         10 & 0.20 & 5875.19 & 2.98 \\
        \hline
         11 & 0.09 & 2378.72 & 0.95 \\
        \hline
        \textbf{Average} & \textbf{0.20} & \textbf{2874.15} & \textbf{6.37} \\
        \hline
    \end{tabular}
\end{table}
\subsection{Detailed Validation on High-Resolution Open Data}

To evaluate the algorithm under high-frequency conditions, we utilized the Almanac of Minutely Power dataset (AMPds) \cite{makonin2016ampds}. This open-source repository captures two years (2012--2014) of minutely natural gas, electricity, and water consumption for a single-family residence in Vancouver, equipped with a natural gas furnace and water heater.

\begin{figure}[t!]
    \centering
    \includegraphics[width=0.95\linewidth]{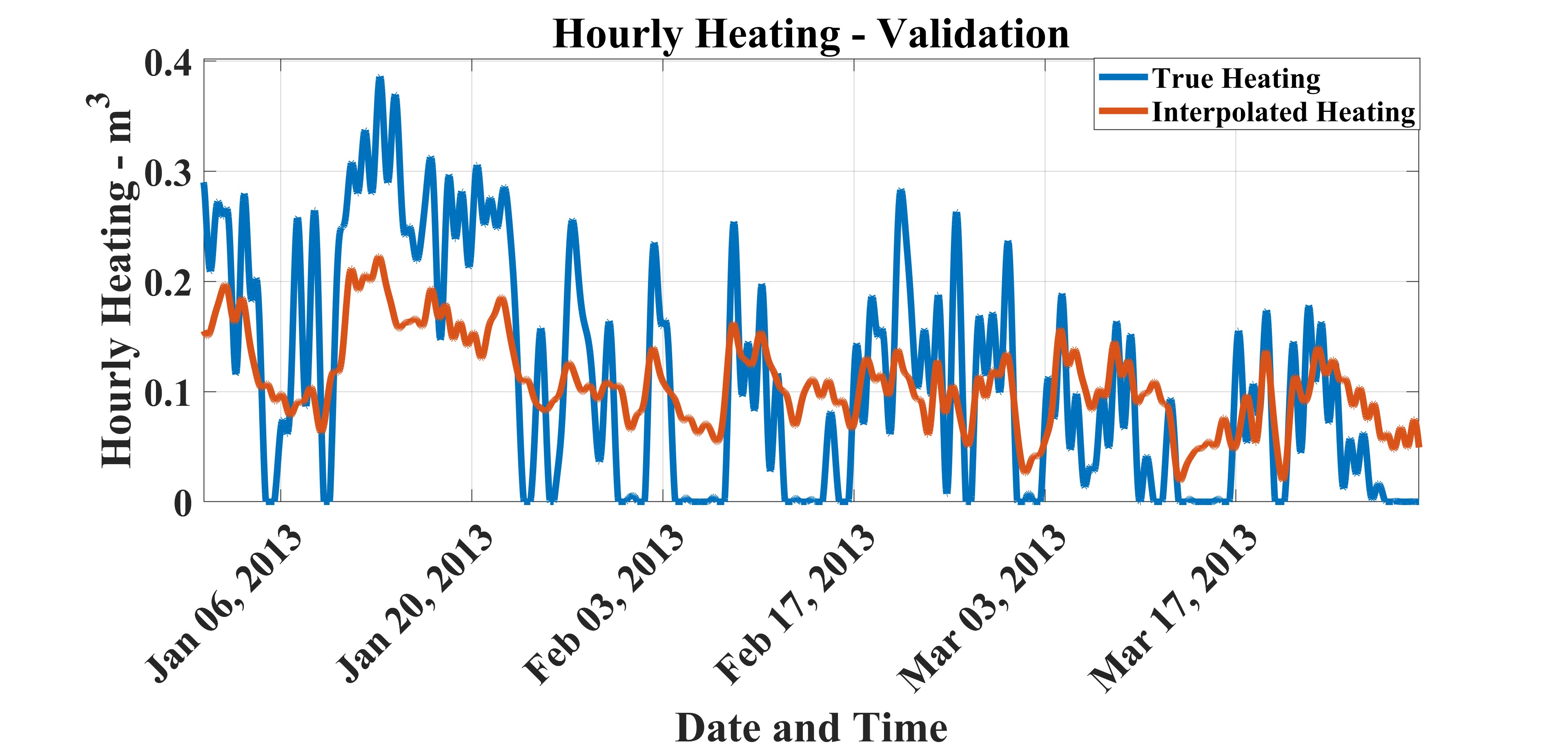}
    \caption{Validation of Heating Component Disaggregation (AMPds). The reconstructed heating profile (red) is compared against the sub-metered ground truth (blue).}
    \label{fig:heating_val}
\end{figure}

\begin{figure}[t!]
    \centering
    \includegraphics[width=0.95\linewidth]{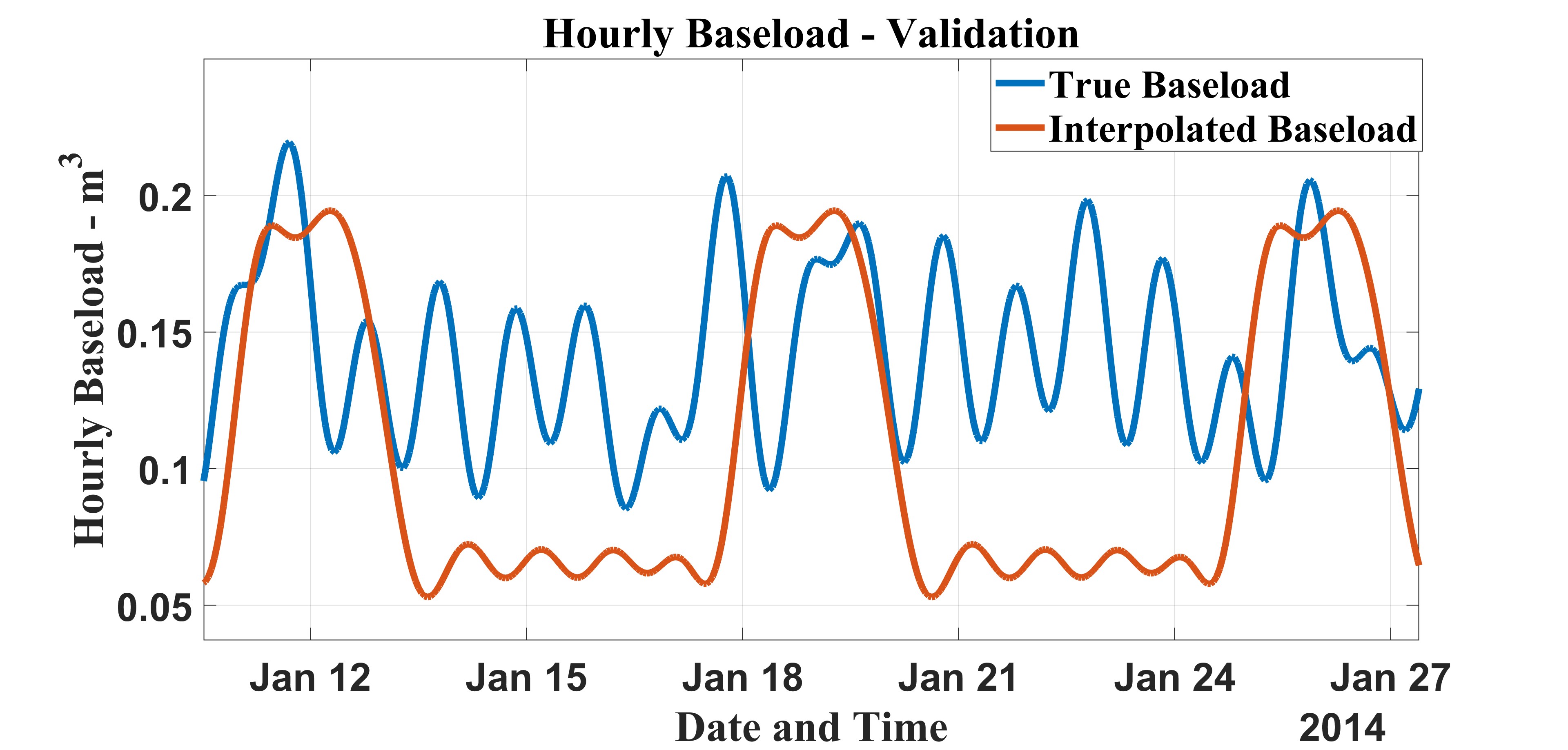}
    \caption{Validation of Baseload Component Disaggregation (AMPds). Comparison of the occupancy-driven baseload prediction (red) versus the actual domestic hot water consumption (blue).}
    \label{fig:baseload_val}
\end{figure}

We applied the proposed framework to this dataset to assess its ability to disaggregate the total gas load into heating and baseload components. Figure \ref{fig:heating_val} and \ref{fig:baseload_val} demonstrate the reconstructed profiles against the ground-truth signals derived from the sub-metered data. The model demonstrated robust performance, achieving a NRMSE of 0.264 and 0.271, and MSPE of 10.1\% and 11.0\% for the domestic hot water (baseload) and heating consumptions, respectively. These results confirm that the dual-stream approach effectively captures the distinct operational dynamics of residential heating systems, even when resolved to minute-level intervals.
\section{Conclusions}\label{sec:conclusions}

This paper presented the \textit{Generalized Proxy-Guided Load and Temporal Disaggregation} framework, a data-driven approach for synthesizing high-resolution thermal profiles from low-frequency billing data. By utilizing hourly electricity as a dynamic behavioral proxy, the method overcomes the data scarcity hindering residential decarbonization. Validation on an 18-month dataset of 11 multi-unit buildings and the high-resolution AMPds repository demonstrated the framework's robustness:(1) Proxy Effectiveness: The normalized electricity proxy ($E_N$) achieved a strong correlation (mean $\{r\}=0.74$) with non-heating gas, significantly outperforming raw electricity (mean $\{r\}=0.28$); (2) Baseload Accuracy: The \textit{Unified Segmentation and Normalized Pooling} strategy effectively mitigated sparsity, yielding a baseload MSPE of 8.58\%; (3) Total Load Fidelity: The dual-stream disaggregation reconstructed total gas consumption with a portfolio-wide MSPE of 6.37\% and NRMSE of 0.20; (4) Component Validation: Testing on the AMPds dataset \cite{makonin2016ampds} confirmed accurate decoupling of heating and baseload components at high resolution, with component-wise MSPEs of approximately 10--11\%.These findings confirm the framework as a scalable, non-intrusive solution for generating granular training data for advanced AI models. Future work will explore integrating additional proxies (e.g., water metering) and extending the pooling strategy to commercial typologies. By enabling precise profiling without sub-metering infrastructure, this research offers a practical tool to accelerate electrification retrofits in the existing building stock.

\bibliographystyle{ieeetr} 
\bibliography{references}  

\end{document}